\documentclass[10pt,conference]{IEEEtran}
\usepackage{mathpazo}
\usepackage{times}

\usepackage{amsmath}
\usepackage{amsfonts}
\usepackage{latexsym}
\usepackage{amssymb}
\usepackage{upref}
\usepackage{theorem}
\usepackage{graphicx}
\usepackage{psfrag}
\usepackage{algorithmic}
\usepackage{algorithm}
\usepackage{verbatim}
\usepackage{cite}
\usepackage{color}

\theoremstyle{plain}
\theorembodyfont{\normalfont\slshape}

\newtheorem{thm}{Theorem$\!$}
\newenvironment{theorem}
{\begin{thm}\hspace*{-1ex}{\bf.}}{\end{thm}}

\newtheorem{lem}{Lemma$\!$}

\newtheorem{prop}{Proposition$\!$}

\newtheorem{cor}{Corollary$\!$}

\newtheorem{defn}{Definition$\!$}

\newtheorem{xmpl}{Example$\!$}
\newenvironment{example}{\begin{xmpl}\hspace*{-1ex}{\bf.}}{\end{xmpl}}

\newtheorem{cnstr}{Construction$\!$}
\newenvironment{construction}{\begin{cnstr}\hspace*{-1ex}{\bf.}}{\end{cnstr}}

\newcounter{enumrom}
\renewcommand{\theenumrom}{(\roman{enumrom})}

\makeatletter
\renewcommand{\@endtheorem}{\endtrivlist}
\makeatother

\makeatletter
\renewcommand{\thefigure}{{\@arabic\c@figure}}
\renewcommand{\fnum@figure}{{\bf Figure\,\thefigure}}
\makeatother

\newcommand{\be}[1]{\begin{equation}\label{#1}}
\newcommand{\ee}{\end{equation}}

\renewcommand{\le}{\leqslant}
\renewcommand{\leq}{\leqslant}

\renewcommand{\geq}{\geqslant}

\newcommand{\Cref}[1]{Co\-ro\-lla\-ry\,\ref{#1}}

\DeclareMathAlphabet{\mathbfsl}{OT1}{cmr}{bx}{it}

\outer\def\proclaim #1. #2\par{\medbreak
 \noindent{\bf#1.\enspace}{\sl#2\par}%
 \ifdim\lastskip<\medskipamount \removelastskip\penalty55\medskip\fi}

\mathchardef\inn="3232
\renewcommand{\in}{{\,\inn\,}}

\begin{document}

\IEEEoverridecommandlockouts 

\title{\Huge\bf Compressed Encoding for\\ Rank Modulation}

\author{
\IEEEauthorblockN{\textbf{Eyal En Gad}}
\IEEEauthorblockA{Electrical~Engineering  \\
California Institute of Technology \\
Pasadena, CA 91125, U.S.A. \\
{\it eengad@caltech.edu}\vspace*{-4.0ex}}
\and
\IEEEauthorblockN{\textbf{Anxiao (Andrew) Jiang}}
\IEEEauthorblockA{Computer Science and Engineering  \\
Texas A\&M University \\
College Station, TX 77843, U.S.A. \\
{\it ajiang@cse.tamu.edu}\vspace*{-4.0ex}} \and \and
\IEEEauthorblockN{\textbf{Jehoshua Bruck}}
\IEEEauthorblockA{Electrical~Engineering  \\
California Institute of Technology \\
Pasadena, CA 91125, U.S.A. \\
{\it bruck@caltech.edu}\vspace*{-4.0ex}}
\thanks{
This work was supported in part by the NSF CAREER Award CCF-0747415, the NSF grant ECCS-0802107, and by an NSF-NRI award.
}
}
\maketitle



\begin{abstract}
Rank modulation has been recently proposed as a scheme for storing information in flash memories. While rank modulation has
advantages in improving write speed and endurance, the current
encoding approach is based on the ``push to the top'' operation
that is not efficient in the general case. We propose a new
encoding procedure where a cell level is raised to be higher than
the minimal necessary subset -instead of all - of the other cell
levels. This new procedure leads to a significantly more
compressed (lower charge levels) encoding.  We derive an upper
bound for a family of codes that utilize the proposed encoding
procedure, and consider code constructions that achieve that bound
for several special cases.
\end{abstract}

\section{Introduction}
\begin{figure*}[hbtp]
\begin{center}
\includegraphics[scale=0.55]{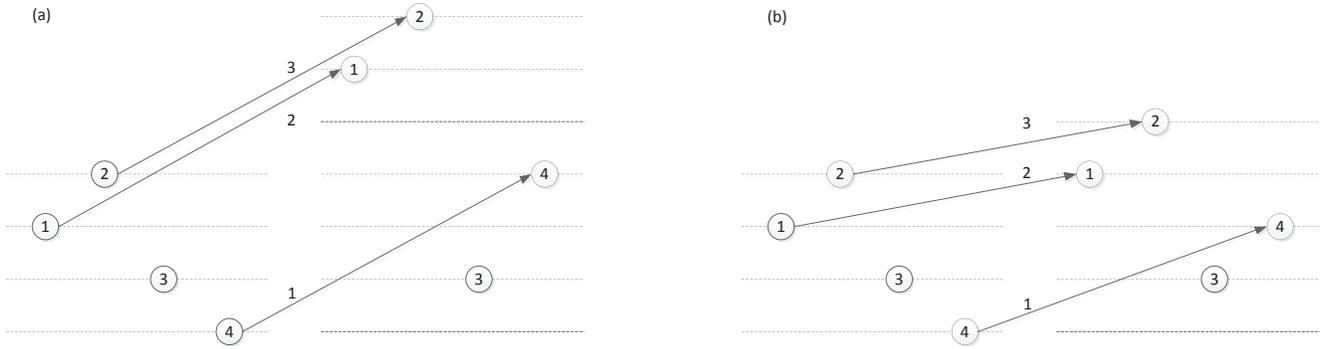}
\caption{\footnotesize
Change the state from $[2,1,3,4]$ to $[2,1,4,3]$ by increasing
cell levels. (a) Using the ``push-to-top'' operations. The labels
beside the three edges show the order of the three ``push-to-top''
operations. (b) Using the more moderate ``minimal-push-up''
operations.} 
\label{fig:transit}
\end{center}
\end{figure*}
With the recent application to flash memories, the rank-modulation
scheme has gained renewed interest as evident in the recent series
of papers
\cite{JiaMatSchBru09,JiaSchBru10,WanJiaBru09,TamSch10,Sch10,EngLanSchBru10}.
Basically, instead of a conventional multi-level cell in which a
symbol in the input alphabet is represented by the charge level of
one cell, in rank modulation, the stored information is
represented by the permutation induced by the relative charge
levels of several cells. The scheme, first described in
\cite{JiaMatSchBru09} in the context of flash memory, works in
conjunction with a simple cell-programming operation called
``push-to-the-top'', which raises the charge level of a single
cell above the rest of the cells. It was suggested in
\cite{JiaMatSchBru09} that this scheme speeds up cell programming
by eliminating the over-programming problem. In addition, it also
reduces corruption due to retention.

The \emph{cost} of changing the state in the scheme -- namely, the
cost of the rewriting step -- is measured by the number of
``push-to-top'' operations that are used, because it represents by
how much the \emph{maximum cell level} among the $n$ cells has
increased~\cite{JiaMatSchBru09}. It is important to minimize this
cell-level increment because the cells have a physical limit that
upper bounds the cell levels. The less the cell levels are
increased, the more rewrites can be performed before a block
erasure operation becomes necessary, and the longer the lifetime
of the memory will be.

We show an example in Fig.~\ref{fig:transit} (a), where the state
of $n=4$ cells needs to change from $\mathbf{u}=[2,1,3,4]$ to
$\mathbf{v}=[2,1,4,3]$. (Here the cells are indexed by
$1,2,\cdots,n$. And their state is denoted by the permutation
$[u_{1},u_{2},\cdots,u_{n}] \in S_{n}$, where cell $u_{1}$ has the
highest charge level and $u_{n}$ has the lowest charge level. For
$i=1,\cdots,n$, cell $u_{i}$ has rank $i$.) Three ``push-to-top''
operations are needed, where cell 4, cell 1 and cell 2 are pushed
sequentially. They are represented by the three edges in the
figure. The \emph{cost} of this rewriting is 3.

We can see from the above example, however, that the
``push-to-top'' operation is a conservative approach. To change
the state from $\mathbf{u}=[2,1,3,4]$ to $\mathbf{v}=[2,1,4,3]$,
when we push cell 4, the level of cell 4 in fact only needs to be
greater than cell 3. There is no need to make it greater than the
levels of all the other $n-1=3$ cells (i.e., cells 1, 2 and 3).
Similarly, when cell 1 is pushed, its level only needs to be
greater than cell 3 and cell 4, instead of cells 2, 3 and 4. So a
more moderate programming approach as shown in
Fig.~\ref{fig:transit} (b) can be taken, and the increment of the
cell levels (in particular, the increment of the maximum cell
level) can be substantially reduced. So the cost of rewriting can
be reduced, which is important for the overall rewriting
performance and the longevity of the memories.

In this work, we consider a programming approach that minimizes
the increase of cell levels. To change the cell state from
$\mathbf{u}=[u_{1},u_{2},\cdots,u_{n}] \in S_{n}$ to
$\mathbf{v}=[v_{1},v_{2},\cdots,v_{n}]\in S_{n}$, we program cells
based on their order in $\mathbf{v}$, so that every cell's level
increases as little as possible:
\begin{itemize}
\item For $i=n-1,n-2,\cdots,1$ do:

\hspace{0mm} $\{$ Increase the level of cell $v_{i}$, to make it
greater than the level of the cell $v_{i+1}$ $\}$.
\end{itemize}
Note that in the above programming process, when cell $v_{i}$ is
programmed, cell $v_{i+1}$ already has the highest level among the
cells $v_{i+1},v_{i+2},\cdots,v_{n}$. We call the programming
operation here the ``minimal-push-up'' operation. (In comparison,
if we program cell $v_{i}$ to make its level greater than the
maximum level among the cells
$v_{1},\cdots,v_{i-1},v_{i+1},\cdots,v_{n}$, then it becomes the
original ``push-to-top'' operation.) The ``minimal-push-up''
approach is robust, too, as it has no risk of overshooting. And it
minimizes increment of the maximum level of the $n$ cells (i.e.,
the rewrite cost).

In section II of the paper we study a discrete model of the rewrite process
that allows us to define the rewrite cost of the
``minimal-push-up'' operation. We show that the cost equals the
maximal rank increase among the elements of the initial
permutation. In section III we turn to the problem of designing rewrite
codes under a worst-case cost constraint. We limit ourselves to
codes that assign a symbol to each one of the states, and show
that such codes for a worst-case cost of 1 cannot achieve a
storage capacity greater than $1-\frac{1}{n}\log_2(\frac{8}{3})$
bits per cell. We consider code constructions that achieve that
bound for $n\leq 5$, where the capacity is $54\%$ higher than that
of previous rank modulation codes. Finally, we show a way to
generalize the construction for higher cost constraints.

\section{Rewrite model and the transition graph}
To design coding schemes, we need a good robust discrete model for
the rewriting. We present here a discrete model for measuring the
rewriting cost, which is suitable for both the ``push-to-top''
approach and the ``minimal-push-up'' approach. To rigorously define the cost of a rewriting step (i.e., a state transition), we will use the
concept of \emph{virtual levels}. Let
$\mathbf{u}=[u_{1},u_{2},\cdots,u_{n}] \in S_{n}$ denote the
current cell state, and let $\mathbf{v}=[v_{1},v_{2},\cdots,v_{n}]
\in S_{n}$ denote the new state that the cells need to change into
via increasing cell levels. Let $d(\mathbf{u}\to \mathbf{v})$
denote the number of push-up operations that are applied to the
cells in order to change the state from $\mathbf{u}$ into
$\mathbf{v}$. For $i=1,2,\cdots,d(\mathbf{u}\to \mathbf{v})$, let
$p_{i}\in [n] \triangleq \{1,2,\cdots,n\}$ denote the integer and
let $B_{i} \subseteq [n]\setminus \{p_{i}\}$ denote the subset,
such that the $i$-th push-up operation is to increase the
$p_{i}$-th cell's level to make it greater than the levels of all
the cells in $B_{i}$. (For example, for the rewriting in
Fig.~\ref{fig:transit} (a), we have $d(\mathbf{u}\to
\mathbf{v})=3$, $p_{1}=4$, $B_{1}=\{1,2,3\}$, $p_{2}=1$,
$B_{2}=\{2,3,4\}$, $p_{3}=2$, $B_{3}=\{1,3,4\}$. And for the
rewriting in Fig.~\ref{fig:transit} (b), we have $d(\mathbf{u}\to
\mathbf{v})=3$, $p_{1}=4$, $B_{1}=\{3\}$, $p_{2}=1$,
$B_{2}=\{3,4\}$, $p_{3}=2$, $B_{3}=\{1,3,4\}$.) Clearly, such
push-up operations have no risk of overshooting.

For the current state $\mathbf{u}$, we assign the \emph{virtual
levels} $n,n-1,\cdots,2,1$ to the cells
$u_{1},u_{2},\dots,u_{n-1},u_{n}$, respectively. The greater a
cell's level is, the greater its \emph{virtual level} is. It
should be noted that when the virtual level increases by one, the
increase in the actual cell level is not a constant because it
depends on the actual programming process, which is noisy.
However, it is reasonable to expect that when we program a cell
$a$ to make its level higher than a cell $b$, the difference
between the two cell levels will concentrate around an expected
value. (For example, a one-shot programming using hot-electron
injection can achieve quite stable programming performance at high
writing speed.) Based on this, we formulate a simple yet robust
discrete model for rewriting, which will be an important tool for
designing coding schemes.

Consider the $i$th push-up operation (for $i= 1,\dots,
d(\mathbf{u}\to \mathbf{v})$), where we increase the level of cell
$p_{i}$ to make it greater than the levels of the cells in
$B_{i}$. For any $j\in [n]$, let $\ell_{j}$ denote cell $j$'s virtual
level before this push-up operation. Then after the
push-up operation, we let the virtual level of cell $p_{i}$ be
\[1+\max_{j\in B_{i}}\ell_{j};\] namely, it is greater than the
maximum virtual level of the cells in $B_{i}$ by one. This
increase represents the increment of the level of cell $p_{i}$.
After the $d(\mathbf{u}\to \mathbf{v})$ push-up operations that
change the state from $\mathbf{u}$ to $\mathbf{v}$, for
$i=1,\dots,n$, let $\ell_{i}'$ denote the virtual level of cell
$i$. We define the \emph{cost} of the rewriting process as the
increase in the maximum virtual level of the $n$ cells, which is
\[\max_{i\in [n]}\ell_{i}' - n = \ell_{v_{1}}' - n.\]

\begin{example}
For the rewriting process shown in Fig.~\ref{fig:transit} (a), the
virtual levels of cells 1, 2, 3, 4 change as $(3,4,2,1)\to
(3,4,2,5)\to (6,4,2,5) \to (6,7,2,5)$. Its cost is 3.

For the rewriting process shown in Fig.~\ref{fig:transit} (b), the
virtual levels of cells 1, 2, 3, 4 change as $(3,4,2,1)\to
(3,4,2,3)\to (4,4,2,3)\to (4,5,2,3)$. Its cost is 1. \hfill $\QED$
\end{example}

\begin{figure}[!t]
\begin{center}
\includegraphics[scale=0.60]{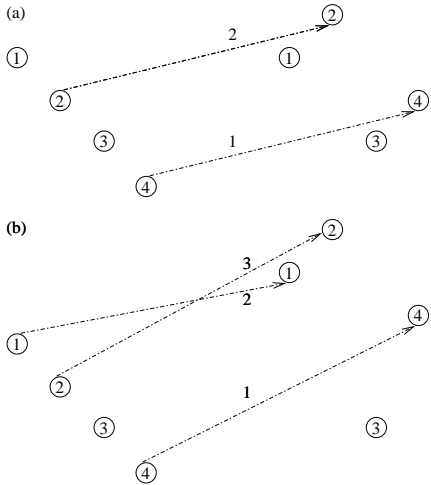}
\caption{\footnotesize
Change the state from $[1,2,3,4]$ to $[2,1,4,3]$. The labels
beside the edges represent the order of the minimal-push-up
operations. (a) The typical programming process, where two cells
are pushed up. (b) In the rare case, if the level of cell 4
exceeds the level of cell 1, then three minimal-push-up operations
are needed.} \label{fig:twoCases}
\end{center}
\end{figure}

We would like to comment further on the robustness of the above
discrete model. The model captures the typical behavior of cell
programming. Yet when the minimal-push-up operations are used, the
number of cells to push in practice is not always a constant when
the old and new states $\mathbf{u},\mathbf{v}$ are given. An
example is shown in Fig.~\ref{fig:twoCases}, where the state needs
to change from $\mathbf{u}=[1,2,3,4]$ to $\mathbf{v}=[2,1,4,3]$.
The typical programming process is shown in
Fig.~\ref{fig:twoCases} (a), where two cells -- cell 4 and then
cell 2 -- are pushed up sequentially. (Note that based on the
discrete model, the rewriting cost is 1. This is consistent with
the increase of the maximum cell level here.) But as shown in
Fig.~\ref{fig:twoCases} (b), in the rare case where cell 4's level
is significantly over-raised to the extent that it exceeds the
level of cell 1, cell 1 will also be programmed, leading to three
minimal-push-up operations in total. However, we would like to
show that above discrete model is still a robust model for the
following reasons. First, in this paper we focus on the typical
(i.e., most probable) behavior of cell programming, where the
rewriting cost matches the actual increase of the maximum cell
level well. In the rare case where cell levels are increased by
too much, additional load balancing techniques over multiple cell
groups can be used to handle it. Second, the rare case -- that a
cell's level is overly increased -- can happen not only with the
minimal-push-up operation but also with the push-to-top operation;
and its effect on the increment of the maximal cell level is
similar for the two approaches. So the discrete model still
provides a fair and robust way to evaluate the rewriting cost of
different state transitions.

\begin{figure}[hbtp]
\begin{center}
\includegraphics[scale=0.65]{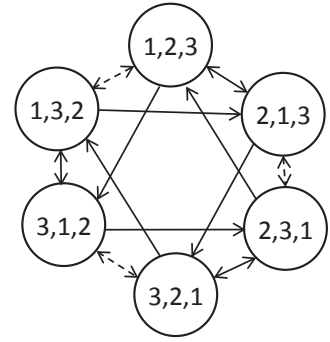}
\caption{\footnotesize
State diagram for the states of three cells, where every edge
represents a state transition of cost 1. The dashed edges are
transitions obtained by ``minimal-push-up'' operations but not
``push-to-top'' operations.}
 \label{fig:transitDiagram}
\end{center}
\end{figure}

In the rest of the paper, we present codes based on state
transitions using the minimal-push-up operations. Given two states
$\mathbf{u}=[u(1),u(2),\cdots,u(n)] \in S_{n}$ and
$\mathbf{v}=[v(1),v(2),\cdots,v(n)] \in S_{n}$, let
$C(\mathbf{u}\to \mathbf{v})$ denote the \emph{cost} of changing
the state from $\mathbf{u}$ to $\mathbf{v}$. (Note that
$u(\cdot),v(\cdot)$ are both functions. Let $u^{-1},v^{-1}$ be
their inverse functions.) The value of $C(\mathbf{u}\to
\mathbf{v})$ can be computed as follows. Corresponding to the old
state $\mathbf{u}$, assign virtual levels $n,n-1,\cdots,1$ to the
cells $u(1),u(2),\cdots,u(n)$, respectively. For $i=1,2,\cdots,n$,
let $\ell_{i}$ denote the virtual level of cell $i$ corresponding
to the new state $\mathbf{v}$. Then based on the programming
process described previously, we can compute
$\ell_{1},\cdots,\ell_{n}$ as follows:
\begin{enumerate}
\item For $i=1,2,\cdots,n$ do:

      $\{$ $\ell_{u(i)} \leftarrow n+1-i$. $\}$

\item For $i=n-1,n-2,\cdots,1$ do:

      $\{$ $\ell_{v(i)} \leftarrow \max\{\ell_{v(i+1)}+1,~\ell_{v(i)}\}$. $\}$
\end{enumerate}
Then we have \[C(\mathbf{u}\to \mathbf{v}) = \ell_{v(1)} - n. \]
It is simple to see that $0 \le C(\mathbf{u}\to \mathbf{v}) \le
n-1$. An example of the rewriting cost is shown in
Fig.~\ref{fig:transitDiagram}.


In the following theorem we present an equivalent definition of the cost. According to the theorem, the cost is equal to the maximal increase in rank among the cells.
\begin{theorem}\label{th:cost}
$$C(\mathbf{u}\to \mathbf{v})=\max_{i\in [n]}(v^{-1}(i)-u^{-1}(i)).$$
\end{theorem}
\begin{IEEEproof}
Assume by induction on k that
$$\ell_{v(k)}=n+1-k+\max_{i\in [k,\dots,n]}(i-u^{-1}(v(i))).$$

In the base case, $k=n$, and
$\ell_{v(n)}=n+1-n+\max_{i\in [n,\dots,n]}(i-u^{-1}(v(i)))=1+n-u^{-1}(v(n)).$
We can see that this is in fact the result of the programming process.
Now we assume that the expression is true for $k$. For $k-1$, by the programming process,
\begin{align*}
\ell_{v(k-1)}
&=\max\{\ell_{v(k)}+1,n+1-u^{-1}(v(k-1)) \} \\
&=\max\{n+1-k+\max_{i\in [k,\dots,n]}(i-u^{-1}(v(i)))+1, \\
&\qquad\qquad n+1-u^{-1}(v(k-1)) \} \\
&\mbox{by the induction assumption} \\
&=n+1-(k-1)+ \\
&\max\{\max_{i\in [k,\dots,n]}(i-u^{-1}(v(i))),k-1-u^{-1}(v(k-1)) \} \\
&=n+1-(k-1)+\max_{i\in [k-1,\dots,n]}(i-u^{-1}(v(i)))
\end{align*}
and the induction is proven.

Now we assign $\ell_{v(1)}$ in the definition of the cost:
\begin{align*}
C(\mathbf{u}\to \mathbf{v}) &= \ell_{v(1)} - n \\
&=n+1-1+\max_{i\in [1,\dots,n]}(i-u^{-1}(v(i)))-n\\
&=\max_{i\in [n]}(v^{-1}(i)-u^{-1}(i))
\end{align*}
\end{IEEEproof}

Codes for rewriting data based on the ``push-to-top'' operation were studied in~\cite{JiaMatSchBru09}.
Since the ``minimal-push-up'' approach has lower rewriting cost than the ``push-to-top'' operation, we can construct rewrite codes with higher rates.

In order to discuss rewriting, we first need to define a decoding
scheme. It is often the case that the alphabet size used by the
user to input data and read stored information differs from the
alphabet size used as internal representation. In our case, data
is stored internally in one of $n!$ different permutations. Let us
assume the user alphabet is $Q = \{1,2,\dots,q\}$. A decoding
scheme is a function $D:S_n\rightarrow Q$ mapping internal states
to symbols from the user alphabet. Suppose the current internal
state is $\mathbf{u}\in S_n$ and the user inputs a new symbol $\alpha \in
Q$. A rewriting operation given $\alpha$ is now defined as moving
from state $\mathbf{u} \in S_n$ to state $\mathbf{v} \in S_n$ such that
$D(\mathbf{v})=\alpha$. The cost of the rewriting operation
is $C(\mathbf{u} \rightarrow \mathbf{v})$.

Next, we define a few terms. Define the transition graph
$G_{n}=(V_{n},A_{n})$ as a directed graph with $V_{n}=S_n$, i.e.,
with $n!$ vertices representing the permutations in $S_n$. There
is a directed edge $\mathbf{u} \rightarrow \mathbf{v}$ if and only
if $C(\mathbf{u} \rightarrow \mathbf{v})=1$. Note that $G_{n}$ is
a regular digraph. Given a vertex $\mathbf{u}\in V_{n}$ and an
integer $r\in \{0,1,\cdots,n-1\}$, we define the ball
$B_{n,r}(\mathbf{u})$ as $B_{n,r}(\mathbf{u})=\{\mathbf{v}\in
V_{n}|C(\mathbf{u} \rightarrow \mathbf{v})\leq r\}$.

\begin{theorem}
\label{lemma:ball_size}
$$|B_{n,r}(\mathbf{u})|=r!(r+1)^{n-r}$$
\end{theorem}
\begin{IEEEproof}
We use induction on $n$. When $n=2$ the statement is trivial. (So
is it when $n=r+1$, where $|B_{r+1,r}(\mathbf{u})|=(r+1)!$.) Now
we assume that the statement is true for $n\le n_{0}$, and
consider $n=n_{0}+1$ and $n>r+1$. Let
$\mathbf{u}=[u(1),u(2),\cdots,u(n)]\in S_{n}$, and without loss
of generality (w.l.o.g.) let $u(1)=n$. Let
$\mathbf{v}=[v(1),v(2),\cdots,v(n)]\in B_{n,r}(\mathbf{u})$.
Let $\hat{\mathbf{u}}=[u(2),u(3),\cdots,u(n)]\in S_{n-1}$, and
let $\hat{\mathbf{v}}\in S_{n-1}$ be obtained from $\mathbf{v}$ by
removing the element $u(1)=n$. By Theorem~\ref{th:cost}, the
first element in $\mathbf{u}$, namely $u(1)=n$, can take one of
the first $r+1$ positions in $\mathbf{v}$. Given that position,
there is a one-to-one mapping between pushing-up the remaining $n-1$
elements from $\mathbf{u}$ to $\mathbf{v}\in S_{n}$ and pushing-up
those $n-1$ elements from $\hat{\mathbf{u}}$ to
$\hat{\mathbf{v}}\in S_{n-1}$, and we have $C(\hat{\mathbf{u}}\to
\hat{\mathbf{v}})=C(\mathbf{u}\to \mathbf{v})\leq r$. So we get
$|B_{n,r}(\mathbf{u})|=(r+1)|B_{n-1,r}(\hat{\mathbf{u}})|=\cdots =
(r+1)^{n-r-1}\cdot (r+1)!=r!(r+1)^{n-r}$.
\end{IEEEproof}

Note that given $\mathbf{u}$, $| \{\mathbf{v}\in S_n| |\mathbf{v}^{-1}(i) - \mathbf{u}^{-1}(i)| \leq r \;\mbox{for}\; 1 \leq i \leq n\}|$
is the size of the ball under infinity norm. When $r=1$, that size is known to be a Fibonacci number \cite{Klo08}.

In addition, we note that $|B_{n,1}(\mathbf{u})|=2^{n-1}$.
Therefore, the out-degree of each vertex in $G_{n}$ is
$2^{n-1}-1$. In comparison, when we allow only the
``push-to-the-top'' operation, $|B_{n,1}(\mathbf{u})|=n$. Hence we
get an exponential increase in the degree, which might lead to an
exponential increase in the rate of rewrite codes. In the next
section we study rewrite codes under a worst-case cost constraint.

\section{Worst-case decoding scheme for rewrite}

In this section, we study codes where the cost of the rewrite
operation is limited by $r$.

\subsection{The case of $n\leq4$}
We start with the case of $r=1$. The first non-trivial case for
$r=1$ is $n=3$. However, for this case the additional
``minimal-push-up'' transitions do not allow for a better rewrite
code. An optimal construction for a graph with only the
``push-to-top'' transitions was described
in~\cite{JiaMatSchBru09}. That construction assigns a symbol to
each state according to the first element in the permutation, for
a total of 3 symbols. It is easy to see that this construction is
also optimal for a graph with the ``minimal-push-up'' transitions.

For greater values of $n$, in order to simplify the construction,
we limit ourselves to codes that assign a symbol to each of the
$n!$ states. We call such codes \emph{full assignment codes}. Note
that better codes for which not all the states are assigned to
symbols might exist. When all of the states are assigned to
symbols, each state must have an edge in $A_{n}$ to at least one
state labelled by each other symbol. We define a set of vertices
$D$ in $G_{n}$ as a \emph{dominating set} if any vertex not in $D$
is the initial vertex of an edge that ends in a vertex in $D$.
Every denominating set is assigned to one symbol. Our goal is
to partition the set of $n!$ vertices into the maximum number of
dominating sets. We start by presenting a construction for $n=4$.

\begin{construction}~\label{con:n4}
Divide the 24 states of $S_4$ into 6 sets of 4 states each, where
each set is a \emph{coset} of $\langle(1,2,3,4)\rangle$, the cyclic \emph{group} generated by $(1,2,3,4)$\footnote{Here $(1,2,3,4)$ is the permutation in
the cycle notation, and $\langle(1,2,3,4)\rangle=\{[1,2,3,4],[2,3,4,1],[3,4,1,2],[4,1,2,3]\}$.}. Map each set to a different symbol.
\end{construction}

\begin{theorem}
Each set in Construction~\ref{con:n4} is a dominating set.
\end{theorem}
\begin{IEEEproof}
Let $I_d$ be the identity permutation, $g=(1,2,3,4)$ and $G=\langle g\rangle$. For each $\mathbf{h}\in S_4$,
$\mathbf{h}G$ is a coset of $G$. For each
$\mathbf{v}=[v(1),\cdots,v(n)]\in \mathbf{h}G$ and each
$\mathbf{u}=[u(1),\cdots,u(n)]\in S_4$ such that $u(1)=v(1)$,
$\mathbf{u}$ has an edge to either $\mathbf{v}$ or $\mathbf{v}*g$.
For example, in the coset $I_dG=G$, for $\mathbf{v}=I_d$ and
$\mathbf{u}\in S_{n}$ such that $u(1)=v(1)=1$, if $u(2)$ is 2 or
3, $\mathbf{u}$ has an edge to $I_d=[1,2,3,4]$, and if $u(2)=4$,
$\mathbf{u}$ has an edge to $I_d*g=[4,1,2,3]$. Since $G$ is a
cyclic group of order 4, for every $\mathbf{u}\in S_4$ there
exists $\mathbf{v}\in \mathbf{h}G$ such that $u(1)=v(1)$, and
therefore $\mathbf{h}G$ is a dominating set.
\end{IEEEproof}

For $k\in [n]$ and $B\subseteq S_n$, we define
$$Pref_k(B)=\{t|\mathbf{s}=tu\;\mbox{for}\; |u|=k\;\mbox{and}\; \mathbf{s}\in B\}$$
where $t,u$ are segments of the permutation $\mathbf{s}$.
For example, $Pref_3(\{[1,2,3,4,5],[1,2,3,5,4],[1,3,2,4,5]\})=\{[1,2],[1,3]\}$.

We provide a lower bound to a dominating set's size.

\begin{theorem}\label{th:domset_bound}
If $D$ is a dominating set of $G_{n}$, then
$$|D|\geq \frac{n!}{\frac{3}{4} \cdot 2^{n-1}}.$$
\end{theorem}
\begin{IEEEproof}
Each $p_3\in Pref_3(S_n)$ is a prefix of 3 different prefixes in
$Pref_2(S_n)$. For example, for $n=5$, $[1,2]$ is a prefix of $\{[1,2,3],[1,2,4],[1,2,5]\}$. Each $\mathbf{v}\in D$ dominates $2^{n-2}$ prefixes in $Pref_2(S_n)$. For example, for $n=4$, every permutation that start with $[1,2],[1,3],[2,1]$ or $[2,3]$ has an edge to $[1,2,3,4]$.
This set of prefixes can be partitioned into
sets of two members, each sharing the same prefix in
$Pref_3(S_n)$. We look at one such set $B_2=\{p_{2,1},p_{2,2}\}$,
and denote by $p_3$ the only member of $Pref_3(B_2)$. Since $D$ is
a dominating set, all of the members of $Pref_2(S_n)$ are
dominated. Therefore, the third prefix $p_{2,3}\notin B_2$ such
that $\{p_3\}=Pref_3(\{B_2,p_{2,3}\})$ is dominated by some
$\mathbf{u}\in D$, $\mathbf{u}\neq \mathbf{v}$. Moreover,
$\mathbf{u}$ dominates also one of the prefixes in $B_2$.
Therefore, at least half of the prefixes in $Pref_2(S_n)$ that
$\mathbf{v}$ dominates are also dominated by at least one other
member of $D$. We denote by $X_{\mathbf{v}}$ the set of prefixes
in $Pref_2(S_n)$ that are dominated by $\mathbf{v}$ and not by any
$\mathbf{u}\neq \mathbf{v}$ such that $\mathbf{u}\in D$, and
denote by $Y_{\mathbf{v}}$ the prefixes in $Pref_2(S_n)$ that are also dominated by at least one such $\mathbf{u}\neq \mathbf{v}$. We further define
$X=\sum_{\mathbf{v}\in D}{|X_{\mathbf{v}}|}$ and
$Y=\sum_{\mathbf{v}\in D}{|Y_{\mathbf{v}}|}$. We have shown that
$|X_{\mathbf{v}}|\leq 2^{n-3}$; so $X\leq 2^{n-3}|D|$. In
addition, we also know that
$|X_{\mathbf{v}}|+|Y_{\mathbf{v}}|=2^{n-2}$, so $X+Y=2^{n-2}|D|$.
By the definition of $Y_{\mathbf{v}}$, we know that $\left
|\bigcup_{\mathbf{v}\in D}{Y_{\mathbf{v}}}\right
|\leq \frac{Y}{2}$, because every element in the above union of sets appears in at least two of the sets. So we get $\frac{n!}{2}=|Pref_2(S_n)| =\left
|\bigcup_{\mathbf{v}\in D}{X_{\mathbf{v}}}\right
|+\left |\bigcup_{\mathbf{v}\in
D}{Y_{\mathbf{v}}}\right | \leq X+\frac{Y}{2} = X
+2^{n-3}|D|-\frac{X}{2} =\frac{X}{2}+2^{n-3}|D| \leq \left
(2^{n-4}+2^{n-3}\right)|D|  =3\cdot 2^{n-4}|D|$. Therefore
$|D|\geq \frac{n!}{3\cdot 2^{n-3}}$.
\end{IEEEproof}

Using the above bound, we can show by a simple calculation that
the rate of any full assignment code $\mathcal{C}$ is $R(\mathcal{C})\leq
1-\frac{1}{n}\log_2{\frac{8}{3}}$ bits per cell. For the case of $n=4$, we see that $|D|\geq 4$. Therefore Construction~\ref{con:n4} is an
\emph{optimal} full assignment code.

\subsection{The case of $n=5$}
In the case of $n=5$, a dominating set consists of at least
$\frac{5!}{3\cdot 2^{5-3}}=10$ members. We present an \emph{optimal} full
assignment code construction with dominating sets of 10 members.
\begin{construction}
\label{con:n5} Divide the 120 states of $S_5$ into 12 sets of 10
states each, where each set is composed of five right cosets of
$\langle(4,5)\rangle$, and two permutations with the same \emph{parity}\footnote{The parity (oddness or evenness) of a permutation $\mathbf{u}$ can be defined as the parity of the number of inversions for $\mathbf{u}$.}
are in the same set if and only if they belong to the same left coset
of $\langle(1,2,4,3,5)\rangle$. Map each set to a different
symbol.
\end{construction}
Let $g_1=(4,5)$ and $g_2=(1,2,4,3,5)$. An example of a dominating set where each row is a right coset of $\langle g_1\rangle$
and each column is a left coset of $\langle g_2\rangle$ is:
\begin{align*}
\{&[1,2,3,4,5],[1,2,3,5,4] \\
&[2,4,5,3,1],[2,4,5,1,3] \\
&[4,3,1,5,2],[4,3,1,2,5] \\
&[3,5,2,1,4],[3,5,2,4,1] \\
&[5,1,4,2,3],[5,1,4,3,2]\}
\end{align*}
\begin{theorem}
Each set $D$ in Construction~\ref{con:n5} is a dominating set.
\end{theorem}
\begin{IEEEproof}
Each right coset of $\langle g_1\rangle$ dominates 4 prefixes in $Pref_3(S_5)$. For example, the coset $\langle g_1\rangle=\{I_d=[1,2,3,4,5],g_1=[1,2,3,5,4]\}$ dominates the prefixes $\{[1,2],[1,3],[2,1],[2,3]\}$. We treat each \emph{coset representative} as a representative of the domination over the 4 prefixes in $Pref_3(S_5)$ that are dominated by the coset. According to the construction, a set of representatives in $D$ that share the same parity is a left coset of $\langle g_2\rangle$. Let one of the cosets of $\langle g_2\rangle$ in $D$ be called $C$. For each $v\in C$, the subset $\{v,v*g_2\}$ represents a domination over the prefix $v(2)$. For example, for $v=I_d$, the subset $\{I_d=[1,2,3,4,5], I_d*g_2=[2,4,5,3,1]\}$ represent a domination over the prefix $[2]$. Since $|\langle g_2\rangle|=5$, $C$ represents a complete domination over $Pref_4(S_5)$, and therefore $D$ is a dominating set.
\end{IEEEproof}

The rate of the code is
$$R=\frac{1}{5}\log_2{12}=0.717\quad\mbox{bits per cell}$$

\subsection{The case of $r\geq 2$}
When the cost constraint is greater than 1, we can generalize the
constructions studied above. We present a construction for the
case $r=n-4$. The construction begins by dividing the $n!$
states $S_n$ into $\frac{n!}{120}$ sets, where two states are
in the same set if and only if their first $n-5$ elements are the
same. The sets are all dominating sets, because we can get to any
set by at most $n-5$ ``push-to-top'' operations. We further divide
each of these sets to 12 sets of 10 members, in the same way as
in Construction~\ref{con:n5}, according to the the last 5 elements of
the permutations. By the properties of construction \ref{con:n5},
each of the smaller sets is still a dominating set. The rate of
the code is $R= \frac{1}{n}\log_2{\frac{n!}{10}}$ bits per cell.

\section{Conclusion}
%

We have presented a programming method that minimizes
rewriting cost for rank modulation, and studied rewrite codes for a worst-case constraint on the cost. The presented codes are optimal full-assignment codes. It remains our future research to extend the code constructions to general code length, non-full assignment codes and average-case cost constraint.



\bibliographystyle{IEEEtranS}
\bibliography{allbib}

\end{document}